\documentstyle[12pt]{article}
\begin{document}

\begin{flushright}
Preprint SSU-HEP-97/04\\
Samara State University
\end{flushright}

\vspace{30mm}

\begin{center}
{\bf CORRECTIONS OF ORDER $(Z\alpha)^6 \frac{m_e^2}{m_\mu}$ IN THE\\
MUONIUM FINE STRUCTURE, INDUCED\\ BY
THREE PHOTON EXCHANGE DIAGRAMS}\footnote{Talk presented at the XII International
Workshop on High Energy Physics and Quantum Field Theory, 4-10 September, 1997,
Samara, Russia}\\
{\bf Faustov R.N.}\\
Scientific Council "Cybernetics" RAS,\\
117333, Vavilova, 40, Moscow, Russia,\\
{\bf Martynenko A.P.}\\
Samara State University, 443011, Ac.Pavlov, 1, Samara, Russia
\end{center}

\def\emline#1#2#3#4#5#6{%
       \put(#1,#2){\special{em:moveto}}%
       \put(#4,#5){\special{em:lineto}}}

\begin{abstract}
In the framework of the quasipotential method we calculate the
contributions of the kind $\alpha^6\ln\alpha$,
$\alpha^6$ of three-photon exchange diagrams to the energy spectrum
of muonium $n^3S_1$ state in the leading order on $m_e/m_\mu$. Analytical
expression of obtained correction for arbitrary principal quantum number
is equal to $(Z\alpha)^6(m_e/m_\mu)m_e(6\ln(2)-35/48)/n^3$.
\end{abstract}

\newpage

The investigation of muonium and positronium fine structure
represents one of the basic tests of quantum electrodynamics, which is
sensitive to radiative corrections of higher order on $\alpha$ \cite{KS}.
Many papers \cite{Ful,Sal,GY,GE} are devoted to the calculation of different
contributions to the fine structure  of hydrogen-like atom energy levels.
The interest to this problem remains unchanged \cite{PG,Y,EG}. The progress,
achieved in the last years during the calculation of logarithmic
contributions of order $\alpha^6 \ln\alpha$ in the positronium fine
structure intervals $(2^3S_1\div 1^3S_1,
2^3S_1\div 2^3P_J)$ \cite{Kh, Fell, FK}, doesn't abolish the necessity of
calculation of higher order corrections $O(\alpha^6)$ \cite{F}. The
development of experimental methods, based on Doppler-free two-photon
spectroscopy, allows the "large" structure intervals to be measured for
the muonium and the positronium \cite{Chu, Chu1, Fee}.
\begin{equation}
\Delta E^{exp.}_{Ps}(2^3S_1\div 1^3S_1)=\left\{{1233607218,9\pm10,7~MHz}
\atop{1233607216,4\pm 3,2~MHz}\right.
\end{equation}
\begin{equation}
\Delta E^{exp.}_{Mu}(2^3S_1\div 1^3S_1)=2455527936\pm 120\pm 140~MHz.
\end{equation}

The increase of the experimental accuracy of muonium fine structure
interval measurements (just as for positronium), planned in the near future, makes very actual
the calculation of radiative corrections of higher order on
$\alpha$. In this paper we have performed studies of the contributions of
order $O(\alpha^6\ln\alpha)$, $O(\alpha^6)$ from three-photon exchange
diagrams to the muonium fine structure. The contribution of these diagrams to
the muonium hyperfine structure was obtained in \cite{BYG}. Our calculations
are based on the Schrodinger-type local quasipotential equation \cite{FM}
\begin{equation}
\left(\frac{b^2}{2\mu_R}-\frac{\vec p^2}{2\mu_R}\right)\psi_M(\vec p)=
\int\frac{d\vec q}{(2\pi)^3}V(\vec p,\vec q,M)\psi_M(\vec q),
\end{equation}
where $b^2=E^2_1-m^2_1=E^2_2-m^2_2,~\mu_R=E_1E_2/M$ is relativistic reduced mass,
$M=E_1+E_2$ is the bound state mass, $m_1, m_2$ are the masses of the electron and
the muon. As an initial approximation of quasipotential $V(\vec p,\vec q,M)$
for the bound state system $(e^-\mu^+)$ we choose the ordinary Coulomb
potential. On the basis of equation (3) in \cite{FM1} we have obtained some
relativistic corrections $m\alpha^6$ in the positronium fine structure from
the one-photon, two-photon interaction and the second order perturbation
theory. There are six diagrams, shown on Fig.1, which determine the three-photon
exchange interaction in the muonium.

\begin{figure}
\unitlength=0.75mm
\special{em:linewidth 1pt}
\linethickness{1pt}
\begin{picture}(150.00,110.00)
\emline{5.00}{5.00}{1}{50.00}{5.00}{2}
\emline{55.00}{5.00}{3}{100.00}{5.00}{4}
\emline{105.00}{5.00}{5}{150.00}{5.00}{6}
\emline{5.00}{50.00}{7}{50.00}{50.00}{8}
\emline{55.00}{50.00}{9}{100.00}{50.00}{10}
\emline{105.00}{50.00}{11}{150.00}{50.00}{12}
\emline{5.00}{65.00}{13}{50.00}{65.00}{14}
\emline{55.00}{65.00}{15}{100.00}{65.00}{16}
\emline{105.00}{65.00}{17}{150.00}{65.00}{18}
\emline{5.00}{110.00}{19}{50.00}{110.00}{20}
\emline{55.00}{110.00}{21}{100.00}{110.00}{22}
\emline{105.00}{110.00}{23}{150.00}{110.00}{24}
\emline{10.00}{110.00}{25}{13.00}{107.00}{26}
\emline{13.00}{107.00}{27}{7.00}{104.00}{28}
\emline{7.00}{104.00}{29}{13.00}{101.00}{30}
\emline{13.00}{101.00}{31}{7.00}{98.00}{32}
\emline{7.00}{98.00}{33}{13.00}{95.00}{34}
\emline{13.00}{95.00}{35}{7.00}{92.00}{36}
\emline{7.00}{92.00}{37}{13.00}{89.00}{38}
\emline{13.00}{89.00}{39}{7.00}{86.00}{40}
\emline{7.00}{86.00}{41}{13.00}{83.00}{42}
\emline{13.00}{83.00}{43}{7.00}{80.00}{44}
\emline{7.00}{80.00}{45}{13.00}{77.00}{46}
\emline{13.00}{77.00}{47}{7.00}{74.00}{48}
\emline{7.00}{74.00}{49}{13.00}{71.00}{50}
\emline{45.00}{110.00}{51}{48.00}{107.00}{52}
\emline{48.00}{107.00}{53}{42.00}{104.00}{54}
\emline{42.00}{104.00}{55}{48.00}{101.00}{56}
\emline{48.00}{101.00}{57}{42.00}{98.00}{58}
\emline{42.00}{98.00}{59}{48.00}{95.00}{60}
\emline{48.00}{95.00}{61}{42.00}{92.00}{62}
\emline{42.00}{92.00}{63}{48.00}{89.00}{64}
\emline{48.00}{89.00}{65}{42.00}{86.00}{66}
\emline{42.00}{86.00}{67}{48.00}{83.00}{68}
\emline{48.00}{83.00}{69}{42.00}{80.00}{70}
\emline{42.00}{80.00}{71}{48.00}{77.00}{72}
\emline{48.00}{77.00}{73}{42.00}{74.00}{74}
\emline{42.00}{74.00}{75}{48.00}{71.00}{76}
\emline{28.00}{110.00}{77}{31.00}{107.00}{78}
\emline{31.00}{107.00}{79}{25.00}{104.00}{80}
\emline{25.00}{104.00}{81}{31.00}{101.00}{82}
\emline{31.00}{101.00}{83}{25.00}{98.00}{84}
\emline{25.00}{98.00}{85}{31.00}{95.00}{86}
\emline{31.00}{95.00}{87}{25.00}{92.00}{88}
\emline{25.00}{92.00}{89}{31.00}{89.00}{90}
\emline{31.00}{89.00}{91}{25.00}{86.00}{92}
\emline{25.00}{86.00}{93}{31.00}{83.00}{94}
\emline{31.00}{83.00}{95}{25.00}{80.00}{96}
\emline{25.00}{80.00}{97}{31.00}{77.00}{98}
\emline{31.00}{77.00}{99}{25.00}{74.00}{100}
\emline{25.00}{74.00}{101}{31.00}{71.00}{102}
\emline{60.00}{110.00}{103}{63.00}{107.00}{104}
\emline{63.00}{107.00}{105}{57.00}{104.00}{106}
\emline{57.00}{104.00}{107}{63.00}{101.00}{108}
\emline{63.00}{101.00}{109}{57.00}{98.00}{110}
\emline{57.00}{98.00}{111}{63.00}{95.00}{112}
\emline{63.00}{95.00}{113}{57.00}{92.00}{114}
\emline{57.00}{92.00}{115}{63.00}{89.00}{116}
\emline{63.00}{89.00}{117}{57.00}{86.00}{118}
\emline{57.00}{86.00}{119}{63.00}{83.00}{120}
\emline{63.00}{83.00}{121}{57.00}{80.00}{122}
\emline{57.00}{80.00}{123}{63.00}{77.00}{124}
\emline{63.00}{77.00}{125}{57.00}{74.00}{126}
\emline{57.00}{74.00}{127}{63.00}{71.00}{128}
\emline{145.00}{110.00}{129}{148.00}{107.00}{130}
\emline{148.00}{107.00}{131}{142.00}{104.00}{132}
\emline{142.00}{104.00}{133}{148.00}{101.00}{134}
\emline{148.00}{101.00}{135}{142.00}{98.00}{136}
\emline{142.00}{98.00}{137}{148.00}{95.00}{138}
\emline{148.00}{95.00}{139}{142.00}{92.00}{140}
\emline{142.00}{92.00}{141}{148.00}{89.00}{142}
\emline{148.00}{89.00}{143}{142.00}{86.00}{144}
\emline{142.00}{86.00}{145}{148.00}{83.00}{146}
\emline{148.00}{83.00}{147}{142.00}{80.00}{148}
\emline{142.00}{80.00}{149}{148.00}{77.00}{150}
\emline{148.00}{77.00}{151}{142.00}{74.00}{152}
\emline{142.00}{74.00}{153}{148.00}{71.00}{154}
\emline{13.00}{71.00}{155}{7.00}{68.00}{156}
\emline{7.00}{68.00}{157}{10.00}{65.00}{158}
\emline{31.00}{71.00}{159}{25.00}{68.00}{160}
\emline{25.00}{68.00}{161}{28.00}{65.00}{162}
\emline{48.00}{71.00}{163}{42.00}{68.00}{164}
\emline{42.00}{68.00}{165}{45.00}{65.00}{166}
\emline{63.00}{71.00}{167}{57.00}{68.00}{168}
\emline{57.00}{68.00}{169}{60.00}{65.00}{170}
\emline{148.00}{71.00}{171}{142.00}{68.00}{172}
\emline{142.00}{68.00}{173}{145.00}{65.00}{174}
\emline{95.00}{110.00}{175}{95.00}{105.00}{176}
\emline{95.00}{105.00}{177}{92.00}{105.00}{178}
\emline{92.00}{105.00}{179}{92.00}{100.00}{180}
\emline{92.00}{100.00}{181}{89.00}{100.00}{182}
\emline{89.00}{100.00}{183}{89.00}{95.00}{184}
\emline{89.00}{95.00}{185}{86.00}{95.00}{186}
\emline{86.00}{95.00}{187}{86.00}{90.00}{188}
\emline{86.00}{90.00}{189}{83.00}{90.00}{190}
\emline{83.00}{90.00}{191}{83.00}{85.00}{192}
\emline{83.00}{85.00}{193}{80.00}{85.00}{194}
\emline{80.00}{85.00}{195}{80.00}{80.00}{196}
\emline{80.00}{80.00}{197}{77.00}{80.00}{198}
\emline{77.00}{80.00}{199}{77.00}{75.00}{200}
\emline{77.00}{75.00}{201}{74.00}{75.00}{202}
\emline{74.00}{75.00}{203}{74.00}{70.00}{204}
\emline{74.00}{70.00}{205}{71.00}{70.00}{206}
\emline{71.00}{70.00}{207}{71.00}{65.00}{208}
\emline{134.00}{110.00}{209}{134.00}{105.00}{210}
\emline{134.00}{105.00}{211}{131.00}{105.00}{212}
\emline{131.00}{105.00}{213}{131.00}{100.00}{214}
\emline{131.00}{100.00}{215}{128.00}{100.00}{216}
\emline{128.00}{100.00}{217}{128.00}{95.00}{218}
\emline{128.00}{95.00}{219}{125.00}{95.00}{220}
\emline{125.00}{95.00}{221}{125.00}{90.00}{222}
\emline{125.00}{90.00}{223}{122.00}{90.00}{224}
\emline{122.00}{90.00}{225}{122.00}{85.00}{226}
\emline{122.00}{85.00}{227}{119.00}{85.00}{228}
\emline{119.00}{85.00}{229}{119.00}{80.00}{230}
\emline{119.00}{80.00}{231}{116.00}{80.00}{232}
\emline{116.00}{80.00}{233}{116.00}{75.00}{234}
\emline{116.00}{75.00}{235}{113.00}{75.00}{236}
\emline{113.00}{75.00}{237}{113.00}{70.00}{238}
\emline{113.00}{70.00}{239}{110.00}{70.00}{240}
\emline{110.00}{70.00}{241}{110.00}{65.00}{242}
\emline{45.00}{50.00}{243}{45.00}{45.00}{244}
\emline{45.00}{45.00}{245}{42.00}{45.00}{246}
\emline{42.00}{45.00}{247}{42.00}{40.00}{248}
\emline{42.00}{40.00}{249}{39.00}{40.00}{250}
\emline{39.00}{40.00}{251}{39.00}{35.00}{252}
\emline{39.00}{35.00}{253}{36.00}{35.00}{254}
\emline{36.00}{35.00}{255}{36.00}{30.00}{256}
\emline{36.00}{30.00}{257}{33.00}{30.00}{258}
\emline{33.00}{30.00}{259}{33.00}{25.00}{260}
\emline{33.00}{25.00}{261}{30.00}{25.00}{262}
\emline{30.00}{25.00}{263}{30.00}{20.00}{264}
\emline{30.00}{20.00}{265}{27.00}{20.00}{266}
\emline{27.00}{20.00}{267}{27.00}{15.00}{268}
\emline{27.00}{15.00}{269}{24.00}{15.00}{270}
\emline{24.00}{15.00}{271}{24.00}{10.00}{272}
\emline{24.00}{10.00}{273}{21.00}{10.00}{274}
\emline{21.00}{10.00}{275}{21.00}{5.00}{276}
\emline{34.00}{50.00}{277}{34.00}{45.00}{278}
\emline{34.00}{45.00}{279}{31.00}{45.00}{280}
\emline{31.00}{45.00}{281}{31.00}{40.00}{282}
\emline{31.00}{40.00}{283}{28.00}{40.00}{284}
\emline{28.00}{40.00}{285}{28.00}{35.00}{286}
\emline{28.00}{35.00}{287}{25.00}{35.00}{288}
\emline{25.00}{35.00}{289}{25.00}{30.00}{290}
\emline{25.00}{30.00}{291}{22.00}{30.00}{292}
\emline{22.00}{30.00}{293}{22.00}{25.00}{294}
\emline{22.00}{25.00}{295}{19.00}{25.00}{296}
\emline{19.00}{25.00}{297}{19.00}{20.00}{298}
\emline{19.00}{20.00}{299}{16.00}{20.00}{300}
\emline{16.00}{20.00}{301}{16.00}{15.00}{302}
\emline{16.00}{15.00}{303}{13.00}{15.00}{304}
\emline{13.00}{15.00}{305}{13.00}{10.00}{306}
\emline{13.00}{10.00}{307}{10.00}{10.00}{308}
\emline{10.00}{10.00}{309}{10.00}{5.00}{310}
\emline{71.00}{110.00}{311}{71.00}{105.00}{312}
\emline{71.00}{105.00}{313}{74.00}{105.00}{314}
\emline{74.00}{105.00}{315}{74.00}{100.00}{316}
\emline{74.00}{100.00}{317}{77.00}{100.00}{318}
\emline{77.00}{100.00}{319}{77.00}{95.00}{320}
\emline{77.00}{95.00}{321}{80.00}{95.00}{322}
\emline{80.00}{95.00}{323}{80.00}{90.00}{324}
\emline{80.00}{90.00}{325}{83.00}{90.00}{326}
\emline{83.00}{90.00}{327}{83.00}{85.00}{328}
\emline{83.00}{85.00}{329}{86.00}{85.00}{330}
\emline{86.00}{85.00}{331}{86.00}{80.00}{332}
\emline{86.00}{80.00}{333}{89.00}{80.00}{334}
\emline{89.00}{80.00}{335}{89.00}{75.00}{336}
\emline{89.00}{75.00}{337}{92.00}{75.00}{338}
\emline{92.00}{75.00}{339}{92.00}{70.00}{340}
\emline{92.00}{70.00}{341}{95.00}{70.00}{342}
\emline{95.00}{70.00}{343}{95.00}{65.00}{344}
\emline{110.00}{110.00}{345}{110.00}{105.00}{346}
\emline{110.00}{105.00}{347}{113.00}{105.00}{348}
\emline{113.00}{105.00}{349}{113.00}{100.00}{350}
\emline{113.00}{100.00}{351}{116.00}{100.00}{352}
\emline{116.00}{100.00}{353}{116.00}{95.00}{354}
\emline{116.00}{95.00}{355}{119.00}{95.00}{356}
\emline{119.00}{95.00}{357}{119.00}{90.00}{358}
\emline{119.00}{90.00}{359}{122.00}{90.00}{360}
\emline{122.00}{90.00}{361}{122.00}{85.00}{362}
\emline{122.00}{85.00}{363}{125.00}{85.00}{364}
\emline{125.00}{85.00}{365}{125.00}{80.00}{366}
\emline{125.00}{80.00}{367}{128.00}{80.00}{368}
\emline{128.00}{80.00}{369}{128.00}{75.00}{370}
\emline{128.00}{75.00}{371}{131.00}{75.00}{372}
\emline{131.00}{75.00}{373}{131.00}{70.00}{374}
\emline{131.00}{70.00}{375}{134.00}{70.00}{376}
\emline{134.00}{70.00}{377}{134.00}{65.00}{378}
\emline{60.00}{50.00}{379}{60.00}{45.00}{380}
\emline{60.00}{45.00}{381}{63.00}{45.00}{382}
\emline{63.00}{45.00}{383}{63.00}{40.00}{384}
\emline{63.00}{40.00}{385}{66.00}{40.00}{386}
\emline{66.00}{40.00}{387}{66.00}{35.00}{388}
\emline{66.00}{35.00}{389}{69.00}{35.00}{390}
\emline{69.00}{35.00}{391}{69.00}{30.00}{392}
\emline{69.00}{30.00}{393}{72.00}{30.00}{394}
\emline{72.00}{30.00}{395}{72.00}{25.00}{396}
\emline{72.00}{25.00}{397}{75.00}{25.00}{398}
\emline{75.00}{25.00}{399}{75.00}{20.00}{400}
\emline{75.00}{20.00}{401}{78.00}{20.00}{402}
\emline{78.00}{20.00}{403}{78.00}{15.00}{404}
\emline{78.00}{15.00}{405}{81.00}{15.00}{406}
\emline{81.00}{15.00}{407}{81.00}{10.00}{408}
\emline{81.00}{10.00}{409}{84.00}{10.00}{410}
\emline{84.00}{10.00}{411}{84.00}{5.00}{412}
\emline{71.00}{50.00}{413}{71.00}{45.00}{414}
\emline{71.00}{45.00}{415}{74.00}{45.00}{416}
\emline{74.00}{45.00}{417}{74.00}{40.00}{418}
\emline{74.00}{40.00}{419}{77.00}{40.00}{420}
\emline{77.00}{40.00}{421}{77.00}{35.00}{422}
\emline{77.00}{35.00}{423}{80.00}{35.00}{424}
\emline{80.00}{35.00}{425}{80.00}{30.00}{426}
\emline{80.00}{30.00}{427}{83.00}{30.00}{428}
\emline{83.00}{30.00}{429}{83.00}{25.00}{430}
\emline{83.00}{25.00}{431}{86.00}{25.00}{432}
\emline{86.00}{25.00}{433}{86.00}{20.00}{434}
\emline{86.00}{20.00}{435}{89.00}{20.00}{436}
\emline{89.00}{20.00}{437}{89.00}{15.00}{438}
\emline{89.00}{15.00}{439}{92.00}{15.00}{440}
\emline{92.00}{15.00}{441}{92.00}{10.00}{442}
\emline{92.00}{10.00}{443}{95.00}{10.00}{444}
\emline{95.00}{10.00}{445}{95.00}{5.00}{446}
\emline{10.00}{50.00}{447}{10.00}{45.00}{448}
\emline{10.00}{45.00}{449}{14.00}{45.00}{450}
\emline{14.00}{45.00}{451}{14.00}{40.00}{452}
\emline{14.00}{40.00}{453}{18.00}{40.00}{454}
\emline{18.00}{40.00}{455}{18.00}{35.00}{456}
\emline{18.00}{35.00}{457}{22.00}{35.00}{458}
\emline{22.00}{35.00}{459}{22.00}{30.00}{460}
\emline{22.00}{30.00}{461}{26.00}{30.00}{462}
\emline{26.00}{30.00}{463}{26.00}{25.00}{464}
\emline{26.00}{25.00}{465}{30.00}{25.00}{466}
\emline{30.00}{25.00}{467}{30.00}{20.00}{468}
\emline{30.00}{20.00}{469}{34.00}{20.00}{470}
\emline{34.00}{20.00}{471}{34.00}{15.00}{472}
\emline{34.00}{15.00}{473}{38.00}{15.00}{474}
\emline{38.00}{15.00}{475}{38.00}{10.00}{476}
\emline{38.00}{10.00}{477}{42.00}{10.00}{478}
\emline{42.00}{10.00}{479}{42.00}{5.00}{480}
\emline{60.00}{5.00}{481}{60.00}{10.00}{482}
\emline{60.00}{10.00}{483}{64.00}{10.00}{484}
\emline{64.00}{10.00}{485}{64.00}{15.00}{486}
\emline{64.00}{15.00}{487}{68.00}{15.00}{488}
\emline{68.00}{15.00}{489}{68.00}{20.00}{490}
\emline{68.00}{20.00}{491}{72.00}{20.00}{492}
\emline{72.00}{20.00}{493}{72.00}{25.00}{494}
\emline{72.00}{25.00}{495}{76.00}{25.00}{496}
\emline{76.00}{25.00}{497}{76.00}{30.00}{498}
\emline{76.00}{30.00}{499}{80.00}{30.00}{500}
\emline{80.00}{30.00}{501}{80.00}{35.00}{502}
\emline{80.00}{35.00}{503}{84.00}{35.00}{504}
\emline{84.00}{35.00}{505}{84.00}{40.00}{506}
\emline{84.00}{40.00}{507}{88.00}{40.00}{508}
\emline{88.00}{40.00}{509}{88.00}{45.00}{510}
\emline{88.00}{45.00}{511}{92.00}{45.00}{512}
\emline{92.00}{45.00}{513}{92.00}{50.00}{514}
\emline{110.00}{50.00}{515}{110.00}{45.00}{516}
\emline{110.00}{45.00}{517}{114.00}{45.00}{518}
\emline{114.00}{45.00}{519}{114.00}{40.00}{520}
\emline{114.00}{40.00}{521}{118.00}{40.00}{522}
\emline{118.00}{40.00}{523}{118.00}{35.00}{524}
\emline{118.00}{35.00}{525}{122.00}{35.00}{526}
\emline{122.00}{35.00}{527}{122.00}{30.00}{528}
\emline{122.00}{30.00}{529}{126.00}{30.00}{530}
\emline{126.00}{30.00}{531}{126.00}{25.00}{532}
\emline{126.00}{25.00}{533}{130.00}{25.00}{534}
\emline{130.00}{25.00}{535}{130.00}{20.00}{536}
\emline{130.00}{20.00}{537}{134.00}{20.00}{538}
\emline{134.00}{20.00}{539}{134.00}{15.00}{540}
\emline{134.00}{15.00}{541}{138.00}{15.00}{542}
\emline{138.00}{15.00}{543}{138.00}{10.00}{544}
\emline{138.00}{10.00}{545}{142.00}{10.00}{546}
\emline{142.00}{10.00}{547}{142.00}{5.00}{548}
\emline{142.00}{50.00}{549}{142.00}{45.00}{550}
\emline{142.00}{45.00}{551}{138.00}{45.00}{552}
\emline{138.00}{45.00}{553}{138.00}{40.00}{554}
\emline{138.00}{40.00}{555}{134.00}{40.00}{556}
\emline{134.00}{40.00}{557}{134.00}{35.00}{558}
\emline{134.00}{35.00}{559}{130.00}{35.00}{560}
\emline{130.00}{35.00}{561}{130.00}{30.00}{562}
\emline{130.00}{30.00}{563}{126.00}{30.00}{564}
\emline{126.00}{30.00}{565}{126.00}{25.00}{566}
\emline{126.00}{25.00}{567}{122.00}{25.00}{568}
\emline{122.00}{25.00}{569}{122.00}{20.00}{570}
\emline{122.00}{20.00}{571}{118.00}{20.00}{572}
\emline{118.00}{20.00}{573}{118.00}{15.00}{574}
\emline{118.00}{15.00}{575}{114.00}{15.00}{576}
\emline{114.00}{15.00}{577}{114.00}{10.00}{578}
\emline{114.00}{10.00}{579}{110.00}{10.00}{580}
\emline{110.00}{10.00}{581}{110.00}{5.00}{582}
\emline{126.00}{50.00}{583}{129.00}{47.00}{584}
\emline{129.00}{47.00}{585}{123.00}{44.00}{586}
\emline{123.00}{44.00}{587}{129.00}{41.00}{588}
\emline{129.00}{41.00}{589}{123.00}{38.00}{590}
\emline{123.00}{38.00}{591}{129.00}{35.00}{592}
\emline{129.00}{35.00}{593}{123.00}{32.00}{594}
\emline{123.00}{32.00}{595}{129.00}{29.00}{596}
\emline{129.00}{29.00}{597}{123.00}{26.00}{598}
\emline{123.00}{26.00}{599}{129.00}{23.00}{600}
\emline{129.00}{23.00}{601}{123.00}{20.00}{602}
\emline{123.00}{20.00}{603}{129.00}{17.00}{604}
\emline{129.00}{17.00}{605}{123.00}{14.00}{606}
\emline{123.00}{14.00}{607}{129.00}{11.00}{608}
\emline{129.00}{11.00}{609}{123.00}{8.00}{610}
\emline{123.00}{8.00}{611}{126.00}{5.00}{612}
\end{picture}
\caption{\bf Feynman diagrams of three-photon exchange interaction in the
system $(e^-\mu^+)$.}
\end{figure}
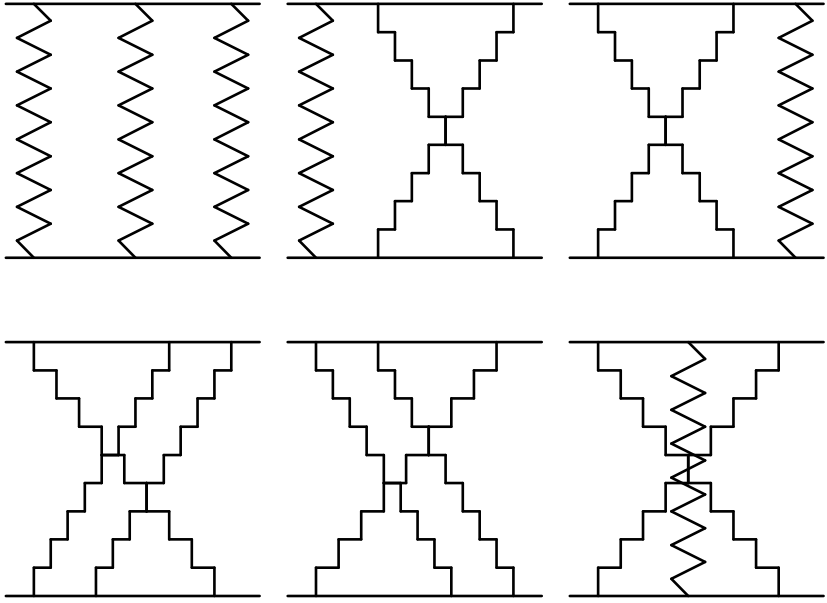

Let consider the first diagram of Fig.1. The corresponding amplitude
already has the factor $\alpha^6$, which appears due to electromagnetic
vertices and Coulomb wave function. So, in the first stage of our
calculations we have neglected by the electron and muon vector momentum
of relative motion in the initial and final states, taking into account
that the necessary accuracy is already achieved. Then the first diagram
amplitude (Fig.1) takes the form:
\begin{equation}
T_1^{3\gamma}=-\frac{(Z\alpha)^3}{4\pi^5}\int d^4p\int d^4p'
\frac{<\gamma_1^\lambda(\hat q_1-\hat p'+m_1)\gamma_1^\nu(\hat p_1-\hat p+m_1)
\gamma_1^\mu>}{(p^2-w^2+i\epsilon)(p'^2-w^2+i\epsilon)[(p-p')^2+i\epsilon]}
\end{equation}
\begin{displaymath}
\frac{<\gamma_2^\mu(\hat p_2+\hat p+m_2)\gamma_2^\nu(\hat q_2+\hat p'+m_2)\gamma_2^\lambda>}
{D_e(p)D_e(p')D_\mu(-p)D_\mu(-p')},
\end{displaymath}
where $D_{e,\mu}(p)$ are the denominators of electron and muon propagators:
\begin{equation}
D(\pm p)=p^2-w^2\pm 2mp^0+i\epsilon,~~~w^2=-b^2,
\end{equation}
and the corner brackets designate the averaging on Dirac bispinors;
$p_1, p_2$ are the four-momenta of particles in the initial state;
$q_1, q_2$ are the particle four-momentum in the final state. As usually is,
the factor $Z\alpha$ emphasizes exchanging character of photon interaction
between particles. The exchange photon propagators were taken in Feynman
covariant gauge. As it famous, using of Coulomb gauge is the most natural
for exchange photons, because the Coulomb interaction dominates in the system
$(e^-\mu^+)$. Nevertheless, the equivalence of Coulomb and Feynman gauges
in the scattering approximation for the three-photon diagrams calculations
was shown in \cite{BYG}. To construct the quasipotential of the system
$(e^-\mu^+)$ with L=0 and J=1, that corresponds to $T^{3\gamma}_1$,
let introduce the projector operator for initial and final states of the
following kind \cite{FM1,Fulton}:
\begin{equation}
\hat\pi=\frac{1}{2\sqrt{2}}\frac{\hat P+M}{M}\hat\varepsilon,
\end{equation}
where $P=p_1+p_2=q_1+q_2$ is full four-momentum of the system,
and $\varepsilon^\mu$ is the vector of $^3S_1$ muonium polarization.
Projecting the particles on $^3S_1$- state by means of (6), we avoid
cumbersome matrix multiplication in the bispinor averages and immediately
pass on to calculation of total trace in (4). As a result, the quasipotential
of first diagram may be written in the form:
\begin{equation}
V_1^{3\gamma}=-\frac{(Z\alpha)^3}{\pi^5}\int d^4p\int d^4p'\frac{F_1(p,p')}
{D_\gamma(p)D_\gamma(p')D_\gamma(p-p')D_e(p)D_e(p')D_\mu(-p)D_\mu(-p')},
\end{equation}
$D_\gamma(p)=p^2-w^2+i\epsilon$,
\begin{equation}
F_1(p,p')=f_{12}(p,p')m_2^2+\frac{1}{3}f_{11}m_2,~~
f_{12}=pp'-4m_1^2-2m_1p_0-2m_1p_0'-2p_0p_0',
\end{equation}
\begin{displaymath}
f_{11}(p,p')=2m_1p'^2+p_0p'^2+10m_1pp'+2p_0pp'+2p_0'pp'+
\end{displaymath}
\begin{displaymath}
2m_1p^2+p_0'p^2+6m_1^2p_0+6m_1^2p_0'+4m_1p_0^2+4m_1p_0'^2-4m_1p_0p_0'.
\end{displaymath}

We kept in (7) only the terms proportional to $m^2_2$ and $m_2$, taking in mind
the determination of contribution to muonium fine structure in the leading
order on parameter $m_1/m_2$. As will soon become evident, we can't restrict
in $F(p,p')$ only by terms $\sim m_2^2$. The quasipotentials of the rest
amplitudes of Fig.1 may be constructed in a similar way. They differ from
each other due to momentum dependence in muonic denominators and to the kind
of functions $f_{i1}$ (i=1,...,6).

The parts of $F_i(p,p')$, proportional to $m_2^2$, coincide in all six
amplitudes. Let remark, that when substitute $\hat\epsilon\rightarrow\gamma_5$
in projector operator (6) ($^1S_0$ state), we obtain the same function
$f_{12}(p,p')$, as for the $^3S_1$ muonium. This means, that the muonium
hyperfine splitting appears as effect of higher order on $m_1/m_2$.
Functions $f_{i1}$ are equal to:
\begin{equation}
f_{21}=-10m_1p'^2-5p_0p'^2+10m_1pp'+4p_0pp'-4p'_0pp'+2m_1p^2+4p'_0p^2+
12p_0m_1^2-
\end{equation}
\begin{displaymath}
-6m_1^2p'_0+4m_1p_0^2+8m_1p_0p'_0-8m_1p'^2_0+4p'_0p_0^2,
\end{displaymath}
\begin{equation}
f_{31}=2m_1^2p'^2+4p_0p'^2+10m_1pp'-4p_0pp'+4p'_0pp'-10m_1p^2-5p'_0p^2-
\end{equation}
\begin{displaymath}
-6m_1^2p_0+12m_1^2p'_0-8m_1p_0^2+8m_1p_0p'_0+4m_1p'^2_0+4p_0p'^2_0,
\end{displaymath}
\begin{equation}
f_{41}=2m_1p'^2+p_0p'^2-2m_1pp'+4p_0pp'+2p'_0pp'-10m_1p^2-8p'_0p^2-12m_1^2p_0+
\end{equation}
\begin{displaymath}
+6m_1^2p'_0+4m_1p_0^2-4m_1p_0p'_0+4m_1p'^2_0-8p'_0p^2_0,
\end{displaymath}
\begin{equation}
f_{51}=-10m_1p'^2-8p_0p'^2-2m_1pp'+2p_0pp'+4p_0'pp'+2m_1p^2+p'_0p^2+6m_1^2p_0-
\end{equation}
\begin{displaymath}
-12m_1^2p'_0+4m_1p^2_0-4m_1p_0p'_0+4m_1p'^2_0-8p'^2_0p_0,
\end{displaymath}
\begin{equation}
f_{61}=-10m_1p'^2-5p_0p'^2-2m_1pp'-4p_0pp'-4p'_0pp'-10m_1p^2-5p'_0p^2-
\end{equation}
\begin{displaymath}
-6m_1^2p_0-6m_1^2p'_0-8m_1p^2_0-4m_1p_0p'_0-8m_1p'^2_0.
\end{displaymath}

Integral function in (4) has simple poles on loop energies $p_0, p'_0$ in
electron, muon and photon propagators.. So, the most natural way of
integration (4) consists in the calculation of integrals on $p_0, p'_0$
at the first step, using the method of residues. But such an approach of
calculation leads, nevertheless, to rather complicated intermediate
expressions, what makes highly questionable its subsequent analytical
integration on spatial momenta $\vec p, \vec p'$. So, we have used different
approach of integration in (4), connected with transformation of muonic
denominators, accounting the necessary calculational accuracy on parameter
$m_1/m_2$. Considering that the spatial momentum of muonic motion in the
intermediate state $|\vec p|< m_2$, we obtain:
\begin{equation}
D_\mu(p)=p^2-w^2+2m_2 p_0\approx 2m_2\left(p_0-\frac{\vec p^2+w^2}{2m_2}+
i\epsilon\right)\approx 2m_2(p_0+i\epsilon),
\end{equation}
where the second approximate equality means that we neglect by the muon
kinetic energy in the intermediate state. Doing so, we suppose that the
integration contour on variable $p_0$ must be closed in the lower halfplane.
Considering the terms, proportional to $m_2^2$ in the numerators of all six
diagrams (function $f_{12}(p,p')$), we have arrived to the need of expression
transformation, which includes the sum of muonic denominators (5). Using the
second approximate equality from (14), we obtain:
\begin{displaymath}
\frac{1}{D_\mu(-p)D_\mu(-p')}+\frac{1}{D_\mu(-p)D_\mu(p'-p)}+\frac{1}{D_\mu(-p')D_\mu(p-p')}+
\end{displaymath}
\begin{displaymath}
+\frac{1}{D_\mu(p)D_\mu(p-p')}+\frac{1}{D_\mu(p')D_\mu(p'-p)}+\frac{1}{D_\mu(p')D_\mu(p)}\approx
\end{displaymath}
\begin{equation}
\approx\frac{(-2\pi i)\delta(p_0)}{2m_2}\frac{(-2\pi i)\delta(p'_0)}
{2m_2}.
\end{equation}

In the energy spectrum the expression (15) will cause the corrections of
order $O(\alpha^4)$, which are canceled by the similar terms from the
iteration part of the quasipotential. Consequently, to find the necessary
contribution of order of $\alpha^6$, we must use first approximate equality
in (14). Taking the difference
\begin{equation}
\frac{1}{2m_2\left(p_0-\frac{\vec p^2+w^2}{2m_2}+i\epsilon\right)}-
\frac{1}{2m_2(p_0+i\epsilon)}\approx\frac{(\vec p^2+w^2)}{4m^2_2(p_0+i\epsilon)^2},
\end{equation}
let represent the quantity $1/D_\mu(p)$ in the form:
\begin{equation}
\frac{1}{D_\mu(p)}\approx\frac{1}{2m_2(p_0+i\epsilon)}+\frac{(\vec p^2+w^2)}
{4m^2_2(p_0+i\epsilon)^2}.
\end{equation}
Second addendum of (17) is of higher order on $m_1/m_2$ in comparison with the
first. But it leads to the necessary order correction  on the other
parameter $\alpha$. Using the splitting (17), in the sum (15), let extract the terms,
which generate the correction $O(\alpha^6)$ and $O(\alpha^6\ln\alpha)$ in
the energy spectrum. We may write them in the following manner:
\begin{equation}
\frac{(\vec p'^2+w^2)}{8m^3_2}\left[\frac{2\pi i\delta(p'_0)}{(p_0+i\epsilon)^2}-
\frac{2\pi i\delta(p'_0-p_0)}{(p_0+i\epsilon)^2}-\frac{2\pi i\delta(p_0)}{(p'_0+
i\epsilon)^2}\right]+
\end{equation}
\begin{displaymath}
+\frac{(\vec p^2+w^2)}{8m^3_2}\left[\frac{-2\pi i\delta(p'_0)}{(p_0+i\epsilon)^2}-
\frac{2\pi i\delta(p'_0-p_0)}{(p_0+i\epsilon)^2}+\frac{2\pi i\delta(p_0)}{(p'_0+
i\epsilon)^2}\right]+
\end{displaymath}
\begin{displaymath}
+\frac{(\vec p-\vec p')^2+w^2}{8m^3_2}\left[-\frac{2\pi i\delta(p'_0)}{(p_0+i\epsilon)^2}+
\frac{2\pi i\delta(p'_0-p_0)}{(p_0+i\epsilon)^2}-\frac{2\pi i\delta(p_0)}{(p'_0+
i\epsilon)^2}\right]
\end{displaymath}

It is evident from three-photon interaction amplitude of the type (6), that
the parts of (18) give the necessary order corrections on $\alpha$ in the studied
fine structure intervals. The same order corrections $O(m_1/m_2)$, as well as (18),
will arise from the quasipotential terms containing the functions
$f_{i1}(p,p')$, when we use the second approximation of (14) for muonic
denominators. To do definite conclusion about the order of appearing terms
in energy spectrum, which are determined by these quasipotential addenda, let
transform them for greater simplification. Let consider for definiteness massless
terms in the function $f_{i1}(p,p')$, proportional to $\sim p^2, p'^2, pp'$:
\begin{equation}
3p^2\left[\frac{1}{D_\mu(p)}+\frac{1}{D_\mu(-p)}-\frac{1}{D_\mu(p-p')}-
\frac{1}{D_\mu(p'-p)}\right]+
\end{equation}
\begin{displaymath}
+3p'^2\left[\frac{1}{D_\mu(-p')}+\frac{1}{D_\mu(p')}-\frac{1}{D_\mu(p'-p)}-
\frac{1}{D_\mu(p-p')}\right]-
\end{displaymath}
\begin{displaymath}
-6pp'\left[\frac{1}{D_\mu(-p')}+\frac{1}{D_\mu(p')}+\frac{1}{D_\mu(-p)}+
\frac{1}{D_\mu(p)}-\frac{1}{D_\mu(p'-p)}-\frac{1}{D_\mu(p-p')}\right]\approx
\end{displaymath}
\begin{displaymath}
\approx\frac{3p^2}{2m_2}\left[-2\pi i\delta(p_0)+2\pi i\delta(p_0-p'_0)\right]+
\frac{3p'^2}{2m_2}\left[-2\pi i\delta(p'_0)+2\pi i\delta(p_0-p'_0)\right]-
\end{displaymath}
\begin{displaymath}
-\frac{6pp'}{2m_2}\left[-2\pi i\delta(p_0)-2\pi i\delta(p'_0)+2\pi i\delta(p_0-p'_0)\right].
\end{displaymath}
The transformation of other terms in functions $f_{i1}(p,p')$ may be carried
out by analogy. The next period of calculation consists in the integration over
four-momenta in expressions (18)-(19). The typical two-loop integral, that
results on this way has the following structure \cite{BYG}:
\begin{equation}
K_i=(4\pi)^2\int\frac{d^4p d^4p'}{-(2\pi)^8}\frac{G_i(p'_0,p_0,m_1)P(\vec p,
\vec p',w)}{(p'^2-w^2+i\epsilon)[(p-p')^2+i\epsilon](p^2-w^2+i\epsilon)D_e(p')
D_e(p)},
\end{equation}
where $G_i(p'_0,p_0,m_1)$ contains one $\delta$ - function, and
$P(\vec p', \vec p, w)$ is a polynom. To calculate fundamental integrals (20)
we have used  Feynman parameterization in order to combine the denominators
of the particle propagators, and the symmetry properties of the integral with
the replacement $p\Leftrightarrow p'$. There are the next set of functions
$G_i(p'_0,p_0,m_1)$, which appear in this paper:
\begin{equation}
G_1=-\frac{2\pi i\delta(p_0-p'_0)2m_1}{(p_0+i\epsilon)^2},~~~
G_2=-\frac{2\pi i\delta(p_0) 2 m_1}{(p'_0+i\epsilon)^2},~~~
G_3=-2\pi i\delta(p_0) 2m_1,
\end{equation}
\begin{displaymath}
G_4=-2\pi i\delta(p'_0)2m_1,~~~G_5=-2\pi i\delta(p_0-p'_0)2m_1.
\end{displaymath}
The results of the integrations for $K_i$ (20) have presented in the table
\cite{BYG}.\\[8mm]
{\bf Table of the integrations $K_i$ (20), appearing in the\\ muonium fine
structure calculations}\\[3mm]
\begin{tabular}{|c|c|c|c|c|} \hline
    & $\vec p^2(\vec p\vec p')$ & $(\vec p\vec p')^2$  & $\vec p'^2(\vec p\vec p')$ & $ w^2(\vec p\vec p') $    \\  \hline
$K_1$ & $2\ln 2-\frac{1}{2}$ & $2\ln 2-\frac{1}{2}$ & $2\ln 2-\frac{1}{2}$ & 0                         \\     \hline
      &$\vec p^2(\vec p'^2-\vec p\vec p')$ & $\vec p\vec p'(\vec p'^2-\vec p\vec p')$ & $w^2(\vec p'^2-\vec p\vec p')$ & $w^2\vec p^2$       \\      \hline
$K_2$ &$\frac{1}{2}\ln\frac{m_1}{2w}-\frac{1}{32}$ &$\frac{1}{4}\ln\frac{m_1}{2w}-\frac{13}{32}$ &$\frac{5}{32}$   & $\frac{2}{3}$            \\      \hline
      &  $\vec p'^2$  & $\vec p\vec p'$   & $\vec p^2 $ &    $w^2$  \\  \hline
$K_3$ &    ---      & $\frac{1}{4}\ln\frac{m_1}{2w}-\frac{1}{4}$  &  $ \frac{1}{2}\ln\frac{m_1}{2w}-\frac{1}{8}$   &   $\frac{1}{8}$  \\  \hline
$K_4$ & $\frac{1}{2}\ln\frac{m_1}{2w}-\frac{1}{8}$ &  $\frac{1}{4}\ln\frac{m_1}{2w}-\frac{1}{4}$ &---   &  $\frac{1}{8}$   \\  \hline
 $K_5$& $\ln 2$ & $\ln 2-\frac{1}{2}$ & $\ln 2$ & 0 \\    \hline
\end{tabular}
\vspace{5mm}

Then the contributions, defined by expressions (8-13) and (18) are correspondingly
equal:
\begin{equation}
\Delta B_1=-\frac{1}{2}(Z\alpha)^6\frac{m_1^2}{m_2}
\end{equation}
\begin{equation}
\Delta B_2=(Z\alpha)^6\frac{m^2_1}{m_2}(6\ln 2-\frac{11}{48})
\end{equation}

We have introduced in (20) the photon mass w to avoid "infrared" singularities.
The "infrared" logarithms $\ln w$, containing this photon mass (see table of
integrals $K_i$), and appearing at intermediate expressions, are mutually
cancelled in the corrections $\Delta B_1$, $\Delta B_2$.

Let consider now the quasipotential addenda, containing the momenta of
particle relative motion in the initial and final states. We denote them by
$\vec r_1$ and $\vec r_2$ correspondingly.
Their consideration leads to modification of $f_{i1}$, which acquire the
following additional terms:
\begin{equation}
\Delta f_{21}=10m_1p'r_2+5p_0p'r_2+m_1pr_2+3p'_0pr_2,
\end{equation}
\begin{equation}
\Delta f_{31}=-m_1p'r_1-3p_0p'r_1-10m_1pr_1-5p'_0pr_1,
\end{equation}
\begin{equation}
\Delta f_{41}=m_1(7p'r_1+5p'r_2+10pr_1+5pr_2)+
6p_0p'r_1+5p_0p'r_2+8p'_0pr_1+5p'_0pr_2,
\end{equation}
\begin{equation}
\Delta f_{51}=m_1(-5p'r_1-10p'r_2-5pr_1-7pr_2)-5p_0p'r_1-8p_0p'r_2-5p'_0pr_1-6p'_0pr_2,
\end{equation}
\begin{equation}
\Delta f_{61}=m_1(-11p'r_1-11p'r_2-11pr_1-11pr_2)-
8p_0p'r_1-8p_0p'r_2-8p'_0pr_1-8p'_0pr_2.
\end{equation}
Using again the symmetry properties of appearing integrals under simultaneous
variable replacement $p\Leftrightarrow p'$, $r_1\Leftrightarrow r_2$, we
obtain cancellation of all integrations in (24)-(28). So, the contribution
the particle relative motion in the fine structure with the accuracy
$O(m_1/m_2)$ is equal to zero. Thus the full value of the calculated
correction $(Z\alpha)^6 m_1^2/m_2$ for hydrogen-like system S-states is
defined as a sum of expressions (22) and (23):
\begin{equation}
\Delta B=(Z\alpha)^6\frac{1}{n^3}\frac{\mu^3}{m_1m_2}\left(6\ln 2-\frac{35}{48}\right).
\end{equation}
Numerical value of obtained contribution (29) for the muonium fine structure
interval $2^3S_1\div 1^3S_1$ takes the value 0,271 ŒHz. We have used in this paper
the diagrammatic approach to the calculation of the corrections of order
$(Z\alpha)^6 m_1^2/m_2$. We have made the most complicated part of
calculations, connected with the two-loop integrations. In order to obtain
the total value of necessary order contribution it is important to complete
these results by similar corrections from one-photon, two-photon interactions
as well as by the results from the second order perturbation theory \cite{FM1}.
It is the aim of future studies.

The work was performed under the financial support of the Russian
Foundation for Fundamental Research (grant no. 96-02-17309).

\end{document}